\documentclass[aps,prc,groupedaddress,showpacs,manuscript]{revtex4}
\usepackage{amssymb}

\usepackage{graphicx}

\begin{document}

\title{Probing the symmetry energy and the degree of isospin equilibrium}

\author {Qingfeng Li$\, ^{1}$\footnote{Fellow of the Alexander von Humboldt Foundation.}
\email[]{Qi.Li@fias.uni-frankfurt.de}, Zhuxia Li$\, ^{2}$, and Horst
St\"{o}cker$\, ^{1,3}$}
\address{
1) Frankfurt Institute for Advanced Studies (FIAS), Johann Wolfgang Goethe-Universit\"{a}t, Max-von-Laue-Str.\ 1, D-60438 Frankfurt am Main, Germany\\
2) China Institute of Atomic Energy, P.O.\ Box 275 (18),
Beijing 102413, P.R.\ China\\
3) Institut f\"{u}r Theoretische Physik, Johann Wolfgang Goethe-Universit\"{a}t, Max-von-Laue-Str.\ 1, D-60438 Frankfurt am Main, Germany\\
 }


\begin{abstract}
The rapidity dependence of the single- and double- neutron to proton
ratios of nucleon emission from isospin-asymmetric but
mass-symmetric reactions Zr+Ru and Ru+Zr at energy range $100
\sim 800$ A MeV and impact parameter range $0\sim 8$ fm is
investigated. The reaction system with isospin-asymmetry and
mass-symmetry has the advantage of simultaneously showing up the
dependence on the symmetry energy and the degree of the isospin
equilibrium. We find that the beam energy- and the impact parameter
dependence of the slope parameter of the double neutron to proton
ratio ($F_D$) as function of rapidity are quite sensitive to the
density dependence of symmetry energy, especially at energies
$E_b\sim 400$ A MeV and reduced impact parameters around $0.5$.
Here the symmetry energy effect  on  the
$F_D$ is enhanced, as compared to the single neutron to proton ratio. The
degree of the equilibrium with respect to isospin (isospin mixing)
in terms of the $F_D$ is addressed and its dependence on the
symmetry energy is also discussed.
\end{abstract}


\pacs{24.10.Lx, 25.75.Dw, 25.75.-q} \maketitle

High energy collision of heavy ions provides a unique tool to study the unbroken
properties of dense nuclear matter at finite temperature, outside of the reach of
conventioned nuclear physics \cite{Stoecker:1986ci,Bass:1998ca}. 
The FOPI Collaboration has recently performed the mixing experiment by
using four mass 96+96 systems $^{96}_{44}$Ru+$^{96}_{44}$Ru,
$^{96}_{40}$Zr+$^{96}_{40}$Zr, $^{96}_{44}$Ru+$^{96}_{40}$Zr, and
$^{96}_{40}$Zr+$^{96}_{44}$Ru to investigate the degrees of the
isospin-equilibrium in the intermediate energy heavy ion
collisions (HICs) \cite{Rami:1999xq,Hong02}. It was found that
complete isospin equilibration is not reached by comparing the
emitted proton number counted in rapidity (see
\cite{Rami:1999xq,Hong02,Hombach:1998rd,Li:2001fg,Gaitanos:2004zh}).
The advantage of taking mixed reactions is that  the isospin
asymmetry shows up more obviously because the effects of the
iso-scalar part of the equation of state (EoS) largely cancel.
This advantage could also be used for probing the density
dependence of the symmetry energy. So far the density dependence
of the nuclear symmetry energy, $E_{sym}(u)$, with $u=\rho/\rho_0$
being the reduced nuclear density, is still largely uncertain, although a
lot of theoretical and experimental efforts have been undertaken
in the past decade \cite{Li:1997px,BaoAnBook01,baranRP}. Recently,
the symmetry energy has been extracted to
$E_{sym}(u)\simeq31.6u^{1.05}$ at subnormal densities  by
comparing with experimental data \cite{Tsang04}  when a free nucleon-nucleon (NN)
elastic cross section has been used \cite{Chen:2004si}, while
$E_{sym}(u)\simeq31.6u^{0.69}$ after considering partly medium
modifications of the nucleon transport \cite{Li:2005jy,Li:2006uc}
were subtracted . The symmetry energy can be investigated by the
double neutron to proton and $\pi^{-}/\pi^{+}$ ratios. The
neutron-rich $^{124,132}{\rm Sn}+^{124}{\rm Sn}$ system and more
isospin-symmetric $^{112}{\rm Sn}+^{112}{\rm Sn}$ system  serve as robust
probes of the nuclear symmetry energy
\cite{Li:2005by,Yong:2005qa}. A very attractive feature of taking
double ratios is the fact that the systematic errors can be
strongly reduced, while the sensitivity of the symmetry energy is
close to the single neutron to proton and $\pi^{-}/\pi^{+}$
ratios.

In the present work, we take two sets of isospin-asymmetric but
mass-symmetric reactions, Zr+Ru and Ru+Zr, to study the symmetry
energy effect on the rapidity dependence of both, single as well
as double unbound neutron to proton ratios. We take advantage of
the mixing reactions to study the influence of the isospin
dependent part of the EoS on the equilibration process. We show
that the slope of the double neutron to proton ratio as
distributed in the rapidity space is --- for mixing reactions ---
closely related to degree of isospin equilibration. The UrQMD
transport model has been updated for simulating intermediate
energy HICs \cite{Bass:1998ca,Weber:2002in,Li:2005zz,Li:2005kq,Li:2005gf,Li:2006ez}. It is
also used for the calculations presented in this work. A soft
equation of state (S-EoS) is adopted \cite{Li:2005gf} (except
otherwise stated). We consider here two sets of symmetry energy
parameterizations: $E_{sym}(u)=34u^{\gamma}$ with $\gamma=0.5$
(soft) and $1.5$ (hard). A non-relativistic medium modification of
the nucleon-nucleon elastic cross section is also considered here
(please see Ref. \cite{Li:2006ez} for details). The method to
construct the freeze-out clusters is the same as presented in our
previous work \cite{Li:2005kq,Li:2006ez}.

\begin{figure}
\includegraphics[angle=0,width=0.8\textwidth]{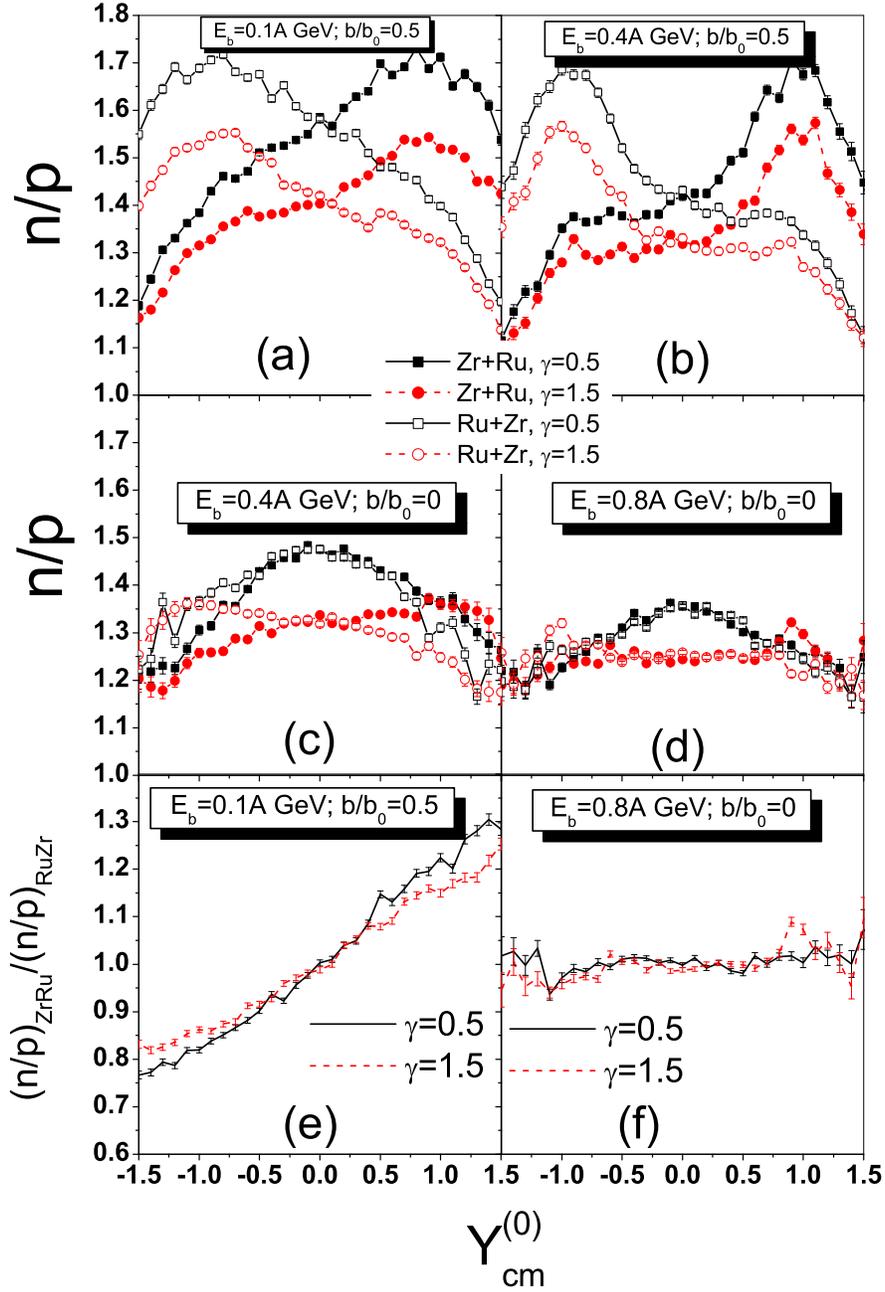}
\caption{(Color online) (a)-(d): Rapidity dependence of neutron to proton ($n/p$)
ratios with symmetry energies of $\gamma=0.5$ and $1.5$. The Zr+Ru
and Ru+Zr reactions with $E_b=100$ A MeV, $b/b_0=0.5$ in plot (a),
$E_b=400$ A MeV, $b/b_0=0.5$ in (b), $E_b=400$ A MeV, $b/b_0=0$ in
(c), and $E_b=800$ A MeV, $b/b_0=0$ in (d) are chosen. (e)-(f):
the corresponding double neutron to proton ratios
($(n/p)_{ZrRu}/(n/p)_{RuZr}$) for (a) and (d) cases (see text).
} \label{fig1}
\end{figure}

We calculate the rapidity ($Y_{cm}^{(0)}=y_{cm}/y_{beam}$ in the
center-of-mass system) dependence of unbound neutrons and protons
in Zr+Ru and Ru+Zr reactions at the beam energies $E_b=100$,
$200$, $400$, $600$, and $800$ A MeV and at the reduced impact
parameters $b/b_0=0$, $0.25$, $0.5$, and $0.75$, where
$b_0=R_{proj}+R_{targ}$ is the maximum impact parameter leading to
nuclear reactions by adopting 'sharp-cutoff' approximation.  The
calculated rapidity dependence of the neutron to proton "$n/p$"
ratios for Zr+Ru and Ru+Zr reactions are shown in Fig.\ \ref{fig1}
(a)-(d) for $E_b=100$ A MeV, $b/b_0=0.5$ in plot (a); $E_b=400$ A
MeV, $b/b_0=0.5$ in (b); $E_b=400$ A MeV, $b/b_0=0$ in (c); and
$E_b=800$ A MeV, $b/b_0=0$ in (d). If equilibrium with respect to
the isospin degree of freedom was reached, the rapidity
distributions of the neutron to proton ratios for the two
different mixing reactions should coincide with each other. Figs.\
\ref{fig1}  (a) and (b) show clearly that  for $b/b_0=0.5$
equilibrium (with respect to isospin degree of freedom) is not
reached, because the centroid of the rapidity dependence of the
neutron to proton ratio is not located at mid-rapidity
\cite{Hombach:1998rd}. For Zr+Ru reactions, the neutron to proton
ratio distribution peaks at the positive rapidity, while for the
inverse Ru+Zr reactions, the distribution peaks at the negative
rapidity. The rapidity dependence of the neutron to proton ratio
is symmetric with respect to $Y_{cm}^{(0)}=0$ for central
reactions at high beam energy (Fig.\ \ref{fig1} (d)), where the
equilibration is nearly reached. It is interesting to see that the
statistics in the UrQMD calculations is perfectly guaranteed since
the calculated results for Zr+Ru and Ru+Zr reactions are
symmetrically distributed very well with respect to
$Y_{cm}^{(0)}=0$, as one expects. This is also a fundamental but
important checkup in the experimental side. As seen for reactions
at $E_b=400$ A MeV and $b/b_0=0$ in Fig.\ \ref{fig1} (c), the
rapidity distribution of the neutron to proton ratio almost
coincides for reactions Zr+Ru and Ru+Zr when the soft-symmetry
energy is chosen, but it deviates strongly when the stiff-symmetry
energy is chosen. This shows the closely relationship between the
effect of the symmetry energy and the equilibration in HICs at
intermediate energies.

Figs.\ \ref{fig1} (e) and (f) present the corresponding rapidity
distributions of double neutron to proton ratio
"$(n/p)_{ZrRu}/(n/p)_{RuZr}$" with the same conditions as Figs.\
\ref{fig1}  (a) and (d), respectively. From Fig.\ \ref{fig1} (e)
it is seen that the double ratio increases almost linearly from
target ($Y_{cm}^{(0)}=-1$) to projectile ($Y_{cm}^{(0)}=1$)
rapidity region for the case of $E_{b}$=100 A MeV and $b/b_0=0.5$.
The two curves corresponding to soft and stiff symmetry energies
cross at the point of rapidity equal to zero,  with positive but
different slopes. And obviously, the slope for $\gamma=0.5$ case
is larger than that for $\gamma=1.5$ case. It is clear that for
isospin asymmetric and mass symmetric reactions Zr+Ru and Ru+Zr,
if the equilibrium with respect to isospin degree of freedom is
totally reached, the double ratio should be unity at all
rapidities. The deviation of $(n/p)_{Zr+Ru}/(n/p)_{Ru+Zr}$ from
unit thus implies the deviation from isospin equilibration. Fig.\
\ref{fig1} (e) shows clearly that the isospin equilibration is not
reached and the choice of symmetry energy influences the isospin
equilibrium process for this case. While for the reactions at
$E_b=800$ A MeV and $b/b_0=0$ presented in Fig.\ \ref{fig1} (f),
the slope is $\sim 0$ and becomes horizontal in large rapidity
region, which means that the isospin equilibration can be roughly
reached. Thus the double neutron to proton ratio as function of
rapidity taking from the mixing reaction system can provide both
the information of density dependence of the symmetry energy and
the degree of isospin equilibrium reached in the reaction.

\begin{figure}
\includegraphics[angle=0,width=0.7\textwidth]{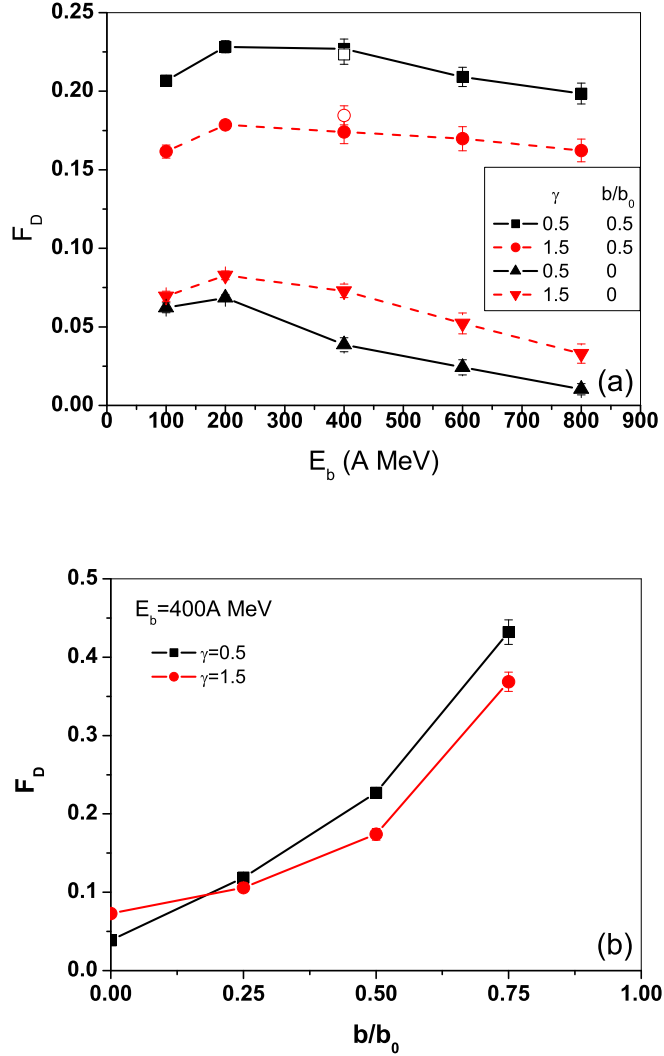}
\caption{(Color online) (a): Beam energy dependence of the slope parameter $F_D$
with symmetry energies with $\gamma=0.5$ and $1.5$, for  impact
parameters $b/b_0=0$ and $0.5$. The $F_D$ values at $E_b=400$ A MeV and at 
$b/b_0=0.5$ with SM-EoS are shown with open dots. (see text)
(b): The centrality dependence of $F_D$ at $E_b=400$ A MeV.}
\label{fig2}
\end{figure}

Further, we study the beam energy and the centrality dependence of
the slope parameters of the double neutron to proton ratio over the rapidity, namely
the $F_{D}$, which is obtained by linearly fitting the rapidity
distribution of the double ratio over $|Y_{cm}^{(0)}|\leq 1$.  Fig.\
\ref{fig2} (a) shows the excitation function of $F_D$ for impact
parameters $b/b_0=0$ and $0.5$ and for symmetry energies 
$\gamma=0.5$ and $1.5$, respectively. The reactions with larger impact parameters show larger
$F_D$ values, which implies that less equilibration is reached.
The $F_D$ values first rise and then drop slightly with increasing
beam energy. It shows a broad maximum at $E_b\sim 200-400$ A MeV  for intermediate impact parameters, while central collisions peak at $200$ A MeV. 
The influence of the symmetry energy on the $F_{D}$ is stronger for $b/b_0=0.5$ than for
$b/b_0=0$. The isoscalar part of the EoS affects
the equilibrium process of the colliding system as well
\cite{Hombach:1998rd}. Fig.\ \ref{fig2} (a) shows $F_D$
at $E_b=400$ A MeV and $b/b_0=0.5$ using a soft EoS with
momentum dependence (SM-EoS, open dots). It is seen that the difference between
the S-EoS and the SM-EoS is very small. This 
indicates that the effect of the isoscalar part of the EoS on 
$F_D$ is largely cancelled, while the effect of the isovector part on
$F_D$ is nicely preserved.

Fig.\ \ref{fig2} (a) also shows that --- for central collisions--- 
$F_D$ as calculated with a stiff symmetry energy is larger than with
a soft one. For $b/b_0=0.5$, the relative magnitude of $F_D$ as 
calculated with the stiff symmetry energy becomes smaller than
that with the soft symmetry energy. The
centrality dependence (with $b/b_0=0$, $0.25$, $0.5$, and $0.75$)
of $F_D$ at beam energy $400$ A MeV is presented in Fig.\ \ref{fig2} (b) . This shows that both curves of $F_D$, 
as calculated with soft and
stiff symmetry energies, increase monotoneously with impact
parameter. The $F_D$ value obtained with the stiff symmetry
energy is larger than that with the soft symmetry energy, when
$b/b_0 \lesssim 0.2$. And they reverse when $b/b_0 \gtrsim 0.2$.
The reversal of the relative magnitudes of $F_{D}$ with stiff or 
soft symmetry energy at small and large impact parameters is due
to the gradual change of the mechanism of particle emission
\cite{Zh05}. Central collisions are more violent,  hence the
compression is stronger than for the collisions with large impact
parameters. Correspondingly, the degree of isospin equilibration is 
both higher and reached more quickly. Nucleons
emitted from central collisions stem mainly from the
mid-rapidity region, where matter is much more
compressed than at the projectile/target region. During the
compression phase, the stiff symmetry potential provides stronger
repulsion (attraction) than the soft one for neutrons (protons) in
the neutron-rich matter. Therefore, the center of collisions is more 
strongly compressed. However, higher pressure caused by the stiff
symmetry potential leads to faster emission of particles. Hence, 
fewer two-body collisions occur and the equilibration decreases.  For
peripheral reaction, the average compression is relatively weak,
Nucleon emission takes place mainly at projectile/target
rapidity, where the nuclear density is low. The low-density
nuclear surface thus plays an important role for nucleon emission.
At low densities, the stiff symmetry potential provides weaker
repulsion (attraction) on neutrons (protons) than the soft one in
the neutron-rich nuclear medium. Higher densities result in the 
projectile/target region, leading to an enhanced number of two-body collisions. The smaller
$F_{D}$ value is evident with the stiff symmetry energy.

The effect of the symmetry energy on $F_D$
becomes the most pronounced at $E_b\sim 400$ A MeV and at
large impact parameters (Fig.\ \ref{fig2}). Compare the isospin effects 
in Fig.\ \ref{fig1} (b) and Fig.\ \ref{fig2} for reactions at
$E_b$ =400 A MeV and $b/b_0=0.5$: The
double neutron to proton ratio exhibits an enhanced sensitivity ($\sim30\%$) to
the density dependence of the symmetry energy as compared to the
single neutron to proton ratio ($\sim10\%$).
Thus, the double neutron to proton ratio (taken from mass
symmetric, but isospin asymmetric mixing systems at energies around
$400$ A MeV and at semi-peripheral impact parameters) is a prominent
probe of the density dependence of the symmetry energy.

In summary, we have
investigated the rapidity dependence of the single- and double
neutron to proton ratios as taken from isospin-asymmetric but
mass-symmetric systems Zr+Ru and Ru+Zr. The beam energy- and the
centrality dependence of the slope parameters of the double
neutron to proton ratio ($F_D$) as function of rapidity are
presented. We find that both, the rapidity distribution of the single
neutron to proton ratio and the $F_D$ are sensitive to the density
dependence of symmetry energy. The study of the beam energy- and
the centrality dependence of the $F_D$ values shows that the
(semi-) peripheral HICs at energies around $E_b\simeq400$ A MeV are
most suited for probing the density dependence of the symmetry
energy. Here the symmetry energy effect is enhanced most strongly 
as is observable in the rapidity dependence of the double neutron to proton
ratio, as compared to the single ratio. 

\section*{Acknowledgments}
Q. Li thanks the Alexander von
Humboldt-Stiftung for a fellowship. This work is partly supported by
the National Natural Science Foundation of China under Grant No.\
10235030 and the Major State Basic Research Development Program of
China under Contract No. G20000774, as well as by GSI, BMBF, DFG,
and Volkswagenstiftung. We gratefully acknowledge support by the Frankfurt Center for
Scientific  Computing (CSC).

\end{document}